\documentclass[aps,reprint,citeautoscript,floatfix,longbibliography]{revtex4-1}
\usepackage{times,revquantum}
\usepackage[utf8]{inputenc}
\usepackage[T1]{fontenc}
\usepackage[capitalize]{cleveref}
\DeclareMathOperator{\sign}{sgn}
\DeclareMathOperator{\pf}{pf}
\newcommand{\up}{{\uparrow}}
\newcommand{\down}{{\downarrow}}
\renewcommand{\Re}{\mathrm{Re}}
\renewcommand{\Im}{\mathrm{Im}}
\renewcommand{\ii}{i}

\begin{document}

\title{Topologically nontrivial Andreev bound states}
\author{Pasquale Marra}
\email{pasquale.marra@keio.jp}
\author{Muneto Nitta}
\email{nitta@phys-h.keio.ac.jp}
\affiliation{Department of Physics, and Research and Education Center for Natural Sciences, Keio University, 4-1-1 Hiyoshi, Yokohama, Kanagawa 223-8521, Japan
}

\begin{abstract}
Andreev bound states are low energy excitations appearing below the particle-hole gap of superconductors, and are expected to be topologically trivial.
Here, we report the theoretical prediction of topologically \emph{nontrivial} Andreev bound states in one-dimensional superconductors.
These states correspond to another topological invariant defined in a synthetic two-dimensional space, the particle-hole Chern number, which we construct in analogy to the spin Chern number in quantum spin Hall systems.
Nontrivial Andreev bound states have distinct features and are topologically nonequivalent to Majorana bound states.
Yet, they can coexist in the same system, have similar spectral signatures, and materialize with the concomitant opening of the particle-hole gap.
The coexistence of Majorana and nontrivial Andreev bound state is the direct consequence of ``double dimensionality'', i.e., the dimensional embedding of the one-dimensional system in a synthetic two-dimensional space, which allows the definition of two distinct topological invariants ($\mathbb{Z}_2$ and $\mathbb{Z}$) in different dimensionalities.
\end{abstract}

\maketitle


Topological phases and their low energy excitations have unprecedented and exotic properties, which partially mimic those of elementary particles in high-energy physics, and may have broad implications for technological applications\cite{hasan_colloquium_2010,qi_topological_2011}.
Prominent examples are the chiral modes and  helical modes realized respectively in quantum Hall\cite{klitzing_new_1980,laughlin_quantized_1981,thouless_quantized_1982} and quantum spin Hall insulators\cite{kane_quantum_2005,kane_$z_2$_2005,sheng_spin_2005,sheng_quantum_2006,bernevig_quantum_2006}, and the Majorana modes in topological superconductors\cite{kitaev_unpaired_2001,alicea_new_2012,leijnse_introduction_2012,beenakker_search_2013,stanescu_majorana_2013,sato_topological_2017,aguado_majorana_2017}.
The connection between topology and low energy excitations is enforced by the bulk-edge correspondence\cite{hatsugai_chern_1993,imura_bulk-edge_2018}, which relates the topological invariants defined in a $d$-dimensional space to the nontrivial modes confined in a lower dimensionality.
Remarkably, topological properties can transcend the spatial dimensions of the physical system, in the sense that the topological invariants may be defined in \emph{synthetic} dimensions\cite{kraus_quasiperiodicity_2016,ozawa_topological_2019}, i.e., additional continuous degrees of freedom which are induced by spatially varying fields in condensed matter\cite{kraus_four-dimensional_2013,park_fractional_2016,thakurathi_fractional_2018}, in particular topological superconductors\cite{kjaergaard_majorana_2012,klinovaja_transition_2012,li_manipulating_2016,marra_controlling_2017}, or can be engineered, e.g., in cold atoms in optical lattices\cite{price_four-dimensional_2015,lohse_exploring_2018,celi_synthetic_2014,marra_fractional_2015,nakajima_topological_2016,lohse_thouless_2016,marra_fractional_2017,zilberberg_photonic_2018}. 
In Josephson junctions, synthetic dimensions can generate nontrivial topological objects such as Weyl points and nonstandard Andreev bound states\cite{riwar_multi-terminal_2016,houzet_majorana-weyl_2019,kotetes_synthetic_2019}.

Additional synthetic dimensions allow the existence of otherwise impossible topological phases.
For example, topological invariants defined in a one-dimensional (1D) space with no symmetries are necessarily trivial\cite{schnyder_classification_2009,kitaev_periodic_2009,ryu_topological_2010}.
Nevertheless, 1D systems such as Thouless quantum pumps exhibit nontrivial phases corresponding to a topological invariant, the Chern number, defined in a synthetic 2D space\cite{thouless_quantization_1983}.
On top of that, there is another possibility:
Synthetic dimensions may allow the coexistence of distinct topological phases characterized by distinct topological invariants, defined in spaces with different dimensions.
Indeed, a $d$-dimensional system embedded in a synthetic $(d+n)$-dimensional space can be described by two distinct topological invariants, e.g., $\mathbb{Z}_2$ and  $\mathbb{Z}$, defined in $d$ and $d+n$ dimensions, and hence occupies two different entries of the periodic table of nontrivial phases\cite{schnyder_classification_2009,kitaev_periodic_2009,ryu_topological_2010}.
Such ``double dimensionality'' may realize, in principle, the coexistence of topologically nonequivalent phases with strikingly different properties.

In this Rapid Communication, we theoretically demonstrate the coexistence of two topologically distinct phases in 1D superconductors due to the dimensional embedding in a 2D synthetic space.
These topological phases correspond to distinct topological invariants:
The familiar Majorana number $M$, defined in the physical 1D space, and the particle-hole (PH) Chern number $C_\mathrm{ph}$, defined in a synthetic 2D space,  which we construct in analogy to the spin Chern number in quantum spin Hall systems\cite{kane_quantum_2005,kane_$z_2$_2005,sheng_spin_2005,sheng_quantum_2006,bernevig_quantum_2006}.
These invariants correspond respectively to Majorana bound states (MBS) and topologically nontrivial Andreev bound states (ABS).
Nontrivial ABS are distinct and topologically nonequivalent to MBS and, unlike trivial ABS\cite{asano_phenomenological_2004,tanaka_theory_2005,golubov_andreev_2009,tanaka_anomalous_2010,liu_zero-bias_2012,kells_near-zero-energy_2012,roy_topologically_2013,stanescu_disentangling_2013,cayao_sns_2015,san-jose_majorana_2016,liu_andreev_2017,liu_distinguishing_2018,moore_two-terminal_2018,moore_quantized_2018,fleckenstein_decaying_2018,awoga_supercurrent_2019}, are fully spin and PH polarized, and protected by PH symmetry.
Moreover, we will show how these distinct topological phases can be realized in a realistic system, i.e., in magnetic atom chains on a conventional superconductor, and we will discuss the differences and similarities between nontrivial ABS and MBS, which are relevant for their experimental realization and identification.

\begin{figure}[t]
\centering
\includegraphics[width=1\columnwidth]{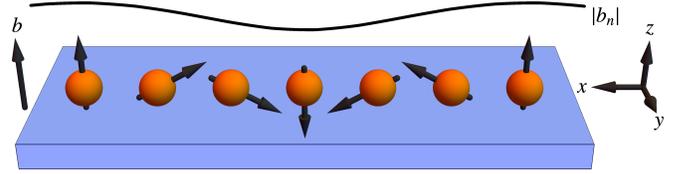}
	\caption{%
A chain of magnetic atoms on the surface of a superconductor.
The amplitude-modulated Zeeman field $\mathbf{b}_n$ is the superposition of an externally applied field $\mathbf{b}$ along the $z$ axis and the field induced by the helical spin order $\Re(e^{\ii(n \theta+\phi)}\boldsymbol{\delta}\mathbf{b})$ in the $zx$ plane.
}
	\label{fig1}
\end{figure}


We thus consider a chain of magnetic atoms\cite{klinovaja_topological_2013,braunecker_interplay_2013,vazifeh_self-organized_2013,pientka_topological_2013,poyhonen_majorana_2014,pientka_unconventional_2014,choy_majorana_2011,nadj-perge_proposal_2013} on the surface of a conventional superconductor, in the presence of a helical spin order\cite{kim_toward_2018} and an externally applied Zeeman field, as in \cref{fig1}.
The essential physics is described by
a Bogoliubov-de~Gennes tight-binding Hamiltonian, which reads
\begin{gather}
H=
\frac12\sum_n
\boldsymbol\Psi_{n}^\dag
\!\cdot\!
\begin{bmatrix}
-\mu \sigma_0
+
\mathbf{b}_n\cdot\boldsymbol{\sigma}
&
\Delta\ii\sigma_y\\
-(\Delta\ii\sigma_y)^*&
\mu \sigma_0
-(\mathbf{b}_n\cdot\boldsymbol{\sigma})^*
\end{bmatrix}
\!\cdot\!
\boldsymbol\Psi_{n}
+\nonumber\\
-
\frac12\sum_n
\boldsymbol\Psi_{n}^\dag
\!\cdot\!
\begin{bmatrix}
t\sigma_0
-\lambda \ii\sigma_y
&
\!\!\!\!\!\!\!\!
0\\
\!\!\!\!
0&
\!\!\!\!\!\!\!\!\!\!
-(t\sigma_0
-\lambda \ii\sigma_y)
\end{bmatrix}
\!\cdot\!
\boldsymbol\Psi_{n+1}
+\text{h.~c.},
\label{eq:Hamiltonian}
\end{gather}
where $\boldsymbol\Psi_n^\dag=[c^\dag_{n\up},c^\dag_{n\down},c_{n\up},c_{n\down}]$ is the Nambu spinor.
Here, $\lambda$ is the intrinsic spin-orbit coupling (SOC) due to the inversion symmetry breaking at the surface, $\Delta$ the spin-singlet superconducting pairing induced by the substrate, and $\mathbf{b}_n$ the total Zeeman field at each site.

The helical spin order is induced by the Ruderman-Kittel-Kasuya-Yosida (RKKY) coupling between localized magnetic moments of the chain, and is resonantly enhanced by the perfect nesting between Fermi momenta $\pm k_\mathrm{F}$ in 1D systems.
This nesting condition fixes the spatial frequency $\theta$ of the helix as $\theta=2k_\mathrm{F}$\cite{braunecker_nuclear_2009,braunecker_spin-selective_2010}, which mandates $\mu=-2t\cos{(\theta/2)}$, which we assume hereafter. 
This assumption is however not essential to the main results\cite{supp}.
Moreover, in the absence of externally applied fields, the spin helix direction is fixed by symmetry.
Indeed, the SOC (along $y$) breaks the $\mathrm{SU}(2)$ spin-rotation symmetry down to ${\mathrm U}(1)$ rotations in the $zx$ plane:
Hence, the helical order becomes pinned to the $zx$ plane for applied fields smaller than the SOC splitting\cite{li_manipulating_2016}.
For simplicity, we assume that the helical order is independent of the externally applied field.
Thus, the total field is
$
\mathbf{b}_n=\mathbf{b}+\Re(e^{\ii(n \theta+\phi)} \boldsymbol{\delta}\mathbf{b})
$,
where $\mathbf{b}$ is the applied field and $\boldsymbol{\delta}\mathbf{b}=(-\ii\delta b,0,\delta b)$.
Here, $\theta$, $\phi$, and $\delta b$ are respectively the spatial frequency, phase-offset, and magnitude of the field induced by the helical spin order.
The cases with $\delta b=0$ and $b=0$ reduce respectively to the well-known regimes where only the uniform\cite{oreg_helical_2010,lutchyn_majorana_2010} and helical fields\cite{choy_majorana_2011,kim_helical_2014,nadj-perge_proposal_2013} are present.
If the applied and helical fields are not perpendicular $\boldsymbol{\delta}\mathbf{b}\cdot \mathbf{b}\neq0$, the total field is amplitude-modulated, $|\mathbf{b}_n|^2=b^2+\delta b^2 + 2\Re(e^{\ii(n\theta+\phi)} \boldsymbol{\delta}\mathbf{b}\cdot \mathbf{b})$, and depends explicitly on the phase-offset $\phi$, which cannot be absorbed by local or global unitary rotations of the spin basis\cite{braunecker_spin-selective_2010,choy_majorana_2011}.
Thus, the energy spectrum and the PH gap depend on the phase-offset $\phi$.
Since we are interested in this regime, we assume that the applied field is coplanar with the helical field.
Besides, one can always rotate the spin basis such that the applied field is parallel to the $z$ axis, which we assume hereafter.
Note that, assuming a rigid and uniformly rotating spin helix, the magnetic order is degenerate in the phase-offset $\phi$, even with external fields $\mathbf{b}\neq \mathbf{0}$, since the coupling between the applied field and the helical order $\propto \mathbf{b} \cdot \mathbf{m}$ vanishes, being the total magnetization $\mathbf{m}=\sum_n\Re(e^{\ii(n \theta+\phi)}\boldsymbol{\delta}\mathbf{b})=\mathbf{0}$.
However, the effect of the applied field on the spin helix may induce a finite magnetization and break the ${\mathrm U}(1)$ invariance.
If these effects are negligible, the phase-offset $\phi$ will become pinned by arbitrarily small local variations of the Zeeman field or by defects and impurities along the chain.
Note also that the SOC induces spin-triplet correlations\cite{heimes_interplay_2015}, which we consider in the Supplemental Material\cite{supp}.

In order to define the topological invariants, it is useful to Fourier-transform the Hamiltonian \eqref{eq:Hamiltonian}, which yields
\begin{gather}
H=
 \frac12
\sum_{k}
\boldsymbol{\Psi}_k^\dag
\!\cdot\!
 \begin{bmatrix}
h(k)
&\Delta \ii\sigma_y
\\
-(\Delta \ii\sigma_y)^*
&
-h(-k)^*
 \end{bmatrix}
\!\cdot\!
 \boldsymbol{\Psi}_k
 +
 \nonumber\\
+\frac14
\sum_{k}
\boldsymbol{\Psi}_{k+\theta}^\dag
\!\cdot\!
 \begin{bmatrix}
e^{\ii \phi}
\boldsymbol{\delta}\mathbf{b}\cdot\boldsymbol{\sigma}
&0 
\\
0&
-
e^{\ii \phi}
\boldsymbol{\delta}\mathbf{b}\cdot\boldsymbol{\sigma}^*
 \end{bmatrix}
\!\cdot\!
 \boldsymbol{\Psi}_k
+
\text{h.~c.},
\label{eq:Hamiltoniank}
\end{gather}
where
$
\boldsymbol\Psi_k^\dag=[c^\dag_{k\up},c^\dag_{k\down},c_{-k\up},c_{-k\down}]
$
and
$h(k)=
\mathbf{b}\!\cdot\!\boldsymbol{\sigma}
-(\mu+2t\cos{k})\sigma_0
+2\lambda\sin{k}\,\sigma_y
$.
We notice that for $\Delta=\lambda=0$, \cref{eq:Hamiltoniank} reduces to the Harper-Hofstadter Hamiltonian realized in topological quantum pumps\cite{harper_single_1955,hofstadter_energy_1976,thouless_quantization_1983,hatsugai_energy_1990}.
Due to the coupling between different momenta, the Hamiltonian is invariant up to momenta translations $k\to k+\theta$.
Assuming a spatial frequency commensurate to the lattice, i.e., $\theta=2\pi p/q$ with $p,q$ integer coprimes, this symmetry induces a periodicity in momentum space $\Delta k=2\pi/q$ and a folding of the energy levels into a reduced Brillouin zone (BZ) $[0,2\pi/q]$.


The model exhibits PH symmetry and broken time-reversal symmetry at any finite field, and belongs to the Altland-Zirnbauer\cite{schnyder_classification_2009,kitaev_periodic_2009,ryu_topological_2010} symmetry class D.
Note that PH symmetry acts as $H(k,\phi)\to-H(-k,\phi)$ in the synthetic BZ (see Supplemental Material\cite{supp}).
Gapped phases in 1D are characterized by a $\mathbb{Z}_2$ topological invariant, the Majorana number\cite{kitaev_unpaired_2001}, defined as $M=\sign(\pf(\ii \widetilde{H}_{0}) \pf(\ii \widetilde{H}_{\pi/q}))$ where  $\widetilde{H}_{k}=\sum_{nm=0}^{q-1} P_{k+n\theta} \widetilde{H} P_{k+m\theta}^\dag$ are the projections of the Hamiltonian $\widetilde{H}$ in the Majorana basis onto the subspace spanned by the momenta $k+n \theta$, with $P_k$ the projector operators, and $k=0,\pi/q$ the time-reversal symmetry points.

\begin{figure}[t]
\centering
\includegraphics[width=1\columnwidth]{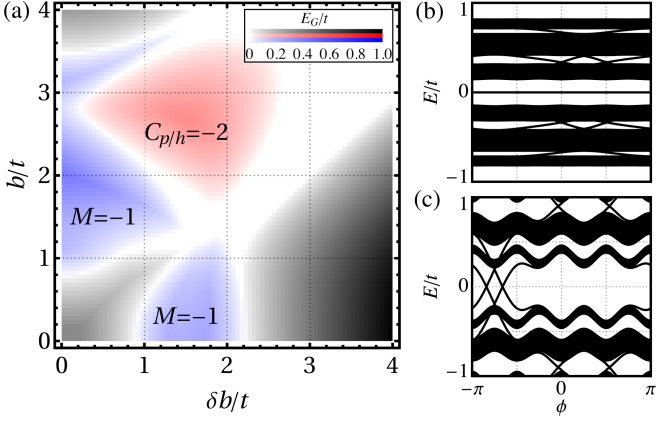}
	\caption{%
(a) Topological phase space for $\theta = \pi/2$ as a function of the helical $\delta b$ and applied fields $b$, and spectra in the nontrivial phases (b-c).
Color intensity is proportional to the global PH gap $E_G$.
Three kinds of globally gapped phases $E_G>0$ are present:
The trivial phase (gray) at small and at very large fields; 
The nontrivial phase with Majorana number $M = -1$ (blue) with (b) MBS at zero energy; 
The nontrivial phase with PH Chern number $C_\mathrm{ph}\neq 0$ (red) with (c) nontrivial ABS below the PH gap, 2 states per edge.
Here, $\mu\approx-1.41 t$, $\Delta=\lambda=t/2$, $\delta b=1.5 t$, $b=0.5 t$ (b), and $b=3 t$ (c).
}
	\label{fig2}
\end{figure}

Phase transitions between trivial and nontrivial phases are determined by the closing of the PH gap, i.e., $E(k,\phi)=0$ for either $k=0$ or $\pi/q$.
For clarity, we will focus here only on the phases which are \emph{globally} gapped, i.e., where the PH gap is finite for any value of the phase-offset $\phi$.
We define the global PH gap as $E_G=\min_{k,\phi} E(k,\phi)$.
Globally gapped phases $E_G>0$ are either trivial or nontrivial.
Conversely, phases which are not globally gapped $E_G=0$, i.e., where the PH gap closes for some values of the phase-offset, may be trivial $M=1$ and nontrivial $M=-1$ depending on the phase-offset $\phi$ (see Ref.~\cite{marra_controlling_2017}).
In \cref{fig2}(a) we plot the value of the global PH gap $E_G$ as a function of the helical field magnitude $\delta b$ and applied field $b$, calculated by direct numerical diagonalization of the Hamiltonian \eqref{eq:Hamiltoniank} for $\theta=\pi/2$ (i.e., $q=4$).
The globally gapped phases $E_G>0$ are separated by domains where the global PH gap vanishes.
We then calculate the Majorana number numerically for each globally gapped phase.
Due to time-reversal symmetry, the phase at zero field $b=\delta b=0$ (and at small fields $b\approx \delta b\approx0$) is obviously trivial.
At larger fields, there are two separated (but topologically equivalent) nontrivial phases with $M=-1$, which are realized respectively for strong applied fields and small (or zero) helical fields $b\gg\delta b$, and for strong helical fields and small (or zero) applied fields $\delta b\gg b$.
The two separated phases with $M=-1$ reduce to the well-known regimes where only a uniform field\cite{oreg_helical_2010,lutchyn_majorana_2010} (with $b>0$ and $\delta b=0$), or the helical field\cite{choy_majorana_2011,nadj-perge_proposal_2013} (with $b=0$ and $\delta {b}>0$) are present.
In these regimes, topological superconductivity is realized respectively for $b^{-}<b<b^{+}$, with $b^{\pm}=[(|\mu|\pm2t)^2+\Delta^2]^{1/2}$ [the $\delta b=0$ axis in \cref{fig1}(a)] and for $b_\mathrm{eff}^{-}<\delta b<b_\mathrm{eff}^{+}$ where $b_\mathrm{eff}^{\pm}=[(|\mu|\pm2t_\mathrm{eff})^2+\Delta^2]^{1/2}$ and 
$t_\mathrm{eff}=t \cos{(\theta/2)}-\lambda \sin{(\theta/2)}$ [the $b=0$ axis in \cref{fig1}(a)], as one can show by unitary rotating \cref{eq:Hamiltonian} and calculating the Majorana number directly.
Nontrivial phases with $M=-1$ exhibits MBS at zero energy, as in \cref{fig2}(b), where we show the energy spectra for $b=\delta b/3$, calculated by direct diagonalization of \cref{eq:Hamiltonian} with open nonperiodic boundary conditions.


\begin{figure*}[t]
\centering
\includegraphics[width=\textwidth]{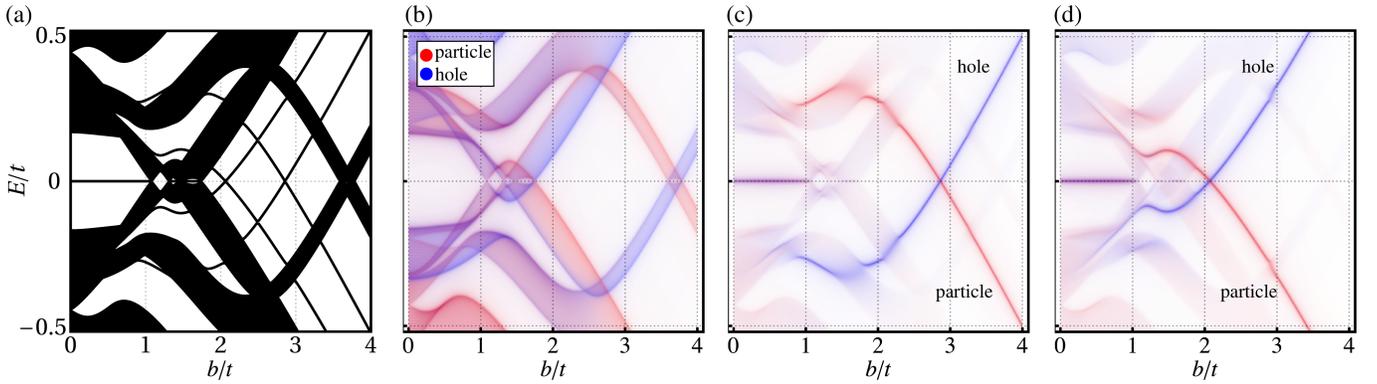}%
	\caption{%
Spectra (a) and LDOS in the bulk (b) and on the left (c) and right (d) edges, as a function of the applied field $b$.
At low applied fields $b\lesssim t$, the model realizes the $M=-1$ nontrivial phase (cf.~\cref{fig1}), with MBS at zero energy localized simultaneously at the left and right edges.
At larger fields, the PH gap closes $E_G=0$ until reaching the nontrivial phase with $C_\mathrm{ph}=-2$ at fields $b\gtrsim t$ (cf.~\cref{fig1}), with 2 PH symmetric and nontrivial ABS localized at each edge of the chain.
The energy of nontrivial ABS at opposite edges is uncorrelated, contrarily to the case of MBS\@.
Moreover, nontrivial ABS are fully PH and spin polarized, whereas MBS are PH symmetric.
}
	\label{fig3}
\end{figure*}

For amplitude-modulated fields $\boldsymbol{\delta}\mathbf{b}\cdot \mathbf{b}\neq0$, the Hamiltonian in \cref{eq:Hamiltoniank} depends periodically on the phase-offset $\phi$, which can be regarded as an additional synthetic (nonspatial) dimension.
The 1D chain is thus embedded in a 2D parameter space, which coincides with a synthetic BZ spanned by the momentum $k\in[0,\pi/q]$ and by the phase-offset $\phi\in[0,2\pi]$.
Topological phases in 2D and symmetry class D are described by a $\mathbb{Z}$ topological invariant.
We notice that the total Chern number is zero due to PH symmetry.
Therefore, to describe the nontrivial globally gapped phases of the model, we shall introduce the PH Chern number, defined as the PH analogue of the spin Chern number\cite{kane_quantum_2005,kane_$z_2$_2005,sheng_spin_2005,sheng_quantum_2006,bernevig_quantum_2006}.
For any globally gapped phase $E_G>0$, which does not close when the superconducting paring is adiabatically turned off $\Delta\to0$, we define
\begin{equation}\label{eq:Chern}
{C}^{\pm}=
\frac{1}{2\pi}
\int_{0}^{2\pi/q} \!\!\! \dd k
\int_{0}^{2\pi}\!\!\! \dd \phi\,
\left[
\Omega_p(k,\phi)
\pm
\Omega_h(k,\phi)
\right]
,
\end{equation}
where $\Omega_{p,h}(k,\phi)=\sum_{i\in {p,h}} 2\Theta(-E_i) \Im \braket{\partial_k \Psi_i | \partial_\phi \Psi_i}$ are the total Berry curvatures in the synthetic BZ, defined respectively for the two PH sectors of the Hamiltonian in \cref{eq:Hamiltoniank} as a sum over all bands with $E_i<0$.
Due to PH symmetry, the total Berry curvatures are $\Omega_h(k,\phi)=-\Omega_p(k,\phi)$, and the Chern number vanishes, $C={C}^+=0$.
The PH Chern number $C_\mathrm{ph}=C^-$ can be nonzero, and it is given by
\begin{equation}
C_\mathrm{ph}=
2\times
\frac{1}{2\pi}
\int_\mathrm{0}^{2\pi/q} \!\!\! \dd k 
\int_{0}^{2\pi}\!\!\! \dd \phi\,
\Omega_p(k,\phi).
\label{eq:ChernPH}
\end{equation}
The PH Chern number is thus an even integer due to PH symmetry.
Notice that the PH Chern number is well-defined only if the phase with $\Delta>0$ can be continuously mapped into a phase with $\Delta=0$, without closing the global PH gap $E_G>0$.
Only in this case indeed, the phase $\Delta>0$ is homeomorphic to the phase $\Delta=0$, where the Hamiltonian becomes block-diagonal in the PH sectors, and the PH Berry curvatures become well-defined.
As a counterexample, notice the PH Chern number $C_\mathrm{ph}$ is not well-defined for the nontrivial phase  $M=-1$, where the gap closes for $\Delta\to0$.

If the helical and applied fields are comparable, the model may realize a nontrivial phase characterized by a nonzero PH Chern number, as shown in \cref{fig2}(a).
Using the Fukui-Hatsugai-Suzuki numerical method\cite{fukui_chern_2005} applied separately in the PH sectors, we find that the PH Chern number of this globally gapped phase is ${C_\mathrm{ph}}=-2$.
The emergence of a nontrivial phase $C_\mathrm{ph}\neq0$ can be understood in terms of a band inversion induced by the applied field.
Considering a continuous transformation $\Delta\to0$, $\lambda\to0$, and $\boldsymbol{\delta}\mathbf{b}=(-\ii \delta b,0,\delta b)\to (0,0,\delta b)$, each of the PH and spin-up and spin-down sectors of the Hamiltonian in \cref{eq:Hamiltoniank} reduce to an Harper-Hofstadter Hamiltonian of spinless electrons on a 1D lattice with harmonic potential $-\mu\pm\delta b \cos{(\theta n+\phi)}$, with $\pm$ for spin up and down, respectively.
Thus, if the transformation does not close the global PH gap, the PH Chern number $C_\mathrm{ph}$ can be obtained as the sum of the corresponding Chern numbers of the Hofstadter butterfly.
Since opposite gaps $\pm\delta b$ of the butterfly have opposite Chern numbers, spin-up and spin-down contributions have opposite signs.
Hence, if we define $j_\up$ and $j_\down$ as the intraband indices of the particle spin-up and spin-down sectors of \cref{eq:Hamiltoniank}, using the diophantine equation characterizing the Hofstadter butterfly Chern numbers, \cref{eq:ChernPH} yields
\begin{equation}
C_\mathrm{ph}=2(C_{j_\up}-C_{j_\down}),
\end{equation}
where
$$
p C_j \equiv j \!\!\! \mod q,
\quad \text{with}\,\,
|C_j|<q/2.
$$
Here, $C_j$ are the Chern numbers labeling each of the intraband gaps $j$ of the Hofstadter butterfly\cite{bellissard_noncommutative_1994,osadchy_hofstadter_2001}.
Since the Hofstadter Chern numbers take all possible integer values $|C_j|< q/2$, the PH Chern number can take all possible even integer values $|C_\mathrm{ph}|<q$.
At zero applied field $b=0$, spin-up and spin-down bands are degenerate, and thus $j_\up=j_\down$, resulting in a trivial phase $C_\mathrm{ph}=0$.
However, spin degeneracy breaks at finite applied fields, and thus bands with $C_{j_\up}\neq C_{j_\down}$ can align at zero energy.
Hence, the band inversion driven by the applied field $b$ can induce a nontrivial phase with $C_\mathrm{ph}\neq0$.

These nontrivial phases correspond to the presence of nontrivial ABS localized at the edges.
Nontrivial ABS are midgap excitations, and are completely PH and spin polarized.
Due to bulk-edge correspondence\cite{hatsugai_chern_1993,imura_bulk-edge_2018}, each edge exhibits a number ${C_\mathrm{ph}}/2$ of particle-like edge states, and the same number of hole-like edge states, which are PH conjugates one of the other.
This has to be contrasted with MBS, which appear as a single zero-energy fermionic state localized at two opposite edges, and which are consequently PH symmetric, i.e., being their own PH conjugates.
\Cref{fig2}(c) shows the energy spectra in the nontrivial phase ${C_\mathrm{ph}}=-2$ (with $b=3 \delta b$) calculated by direct diagonalization of \cref{eq:Hamiltonian} with open nonperiodic boundary conditions.
The spectra show $2\times2$ PH-symmetric, nontrivial ABS inside the PH gap, with 2 edge states for each boundary of the chain.
These ABS are protected by PH symmetry, and robust against perturbations which do not close the gap and do not break PH symmetry (see Supplemental Material\cite{supp}).


Despite the fundamental difference between MBS and nontrivial ABS, there are some similarities that need to be emphasized.
Nontrivial ABS are midgap excitations, and can have zero energy only for fine-tuned values $\phi^*$ of the phase-offset.
However, their energies can be lower than the experimental resolution, and thus the resulting near-zero bias peak can be erroneously attributed to MBS\@.
Most importantly, being topologically protected, they can materialize only concomitantly with the closing and reopening of the PH gap.
Hence, the simultaneous probe of bulk and edge conductance, with the observation of the closing of the gap accompanied by the emergence of a zero-bias peak at the edges\cite{grivnin_concomitant_2019}, cannot be considered as conclusive evidence of MBS\@.
However, nontrivial ABS do not necessarily appear simultaneously with the same energy at the two opposite edges of the nontrivial phase, contrarily to the case of MBS\@.
In order to highlight the differences and similarities between nontrivial ABS and MBS, we show in \cref{fig3} the spectra and the local density of states (LDOS) as a function of the applied field $b$ through the two nontrivial phases $M=-1$ and $C_\mathrm{ph}=-2$, calculated as $\rho_n(E)=-\Im \bra{n}G(E)\ket{n}/\pi$ with $G(E)$ the unperturbed Green's function.
As shown, both the $M=-1$ and the $C_\mathrm{ph}=-2$ nontrivial phases are realized, respectively at low $b\lesssim t$ and large applied fields $b\gtrsim t$, respectively with MBS and nontrivial ABS localized at the edges.
The closing and reopening of the global PH gap coincides with the appearance of nontrivial ABS\@.
Notice that, contrarily to the case of MBS, the energies of the 2 nontrivial ABS at the opposite edges are uncorrelated.
Moreover, whereas MBS have equal spectral weights in the PH sectors (they are PH symmetric), nontrivial ABS are completely PH and spin polarized.


In summary, we found that in the presence of amplitude-modulated fields, a 1D superconductor may exhibits two distinct kinds of nontrivial phases corresponding to two distinct topological invariants, i.e., the Majorana number and the PH Chern number, defined respectively in the 1D and in a synthetic 2D BZ\@.
These nontrivial phases exhibits two distinct kind of edge states, i.e., MBS and nontrivial ABS, with remarkably different properties.
However, their similarities may hinder the detection of Majorana states in magnetic atom chains, in particular in the regime of large applied fields.

This work opens several directions for future research. 
First, nontrivial ABS can be realized in nanowires with amplitude-modulated fields, induced by, e.g., arrays of nanomagnets\cite{kjaergaard_majorana_2012,klinovaja_transition_2012,maurer_designing_2018} or magnetic film substrates in the stripe phase\cite{mohanta_electrical_2019,zhou_tunable_2019},
where they may exhibit distinctive signatures, e.g., in the differential conductance and the Josephson current. 
Moreover, in cold atoms, this model may realize a PH Thouless pump and the direct manipulation of the PH degree of freedom, analogously to the electron spin in spintronics.
Finally, the concept of double dimensionality and of the coexistence of different topological phases can be extended to other contiguous entries of the periodic table of topological phases, or to higher-order topological insulators\cite{schindler_higher-order_2018}.

We thank Daisuke Inotani for useful discussions and suggestions, and the anonymous reviewers whose comments have greatly improved this manuscript.
This work is supported by the Ministry of Education, Culture, Sports, Science, and Technology (MEXT)-Supported Program for the Strategic Research Foundation at Private Universities ``Topological Science'' (Grant No.~S1511006).
The work of M.N. is also supported in part by the Japan Society for the Promotion of Science (JSPS) Grant-in-Aid for Scientific Research (KAKENHI) Grants No. 16H03984 and No. 18H01217 and by a
Grant-in-Aid for Scientific Research on Innovative Areas ``Topological Materials Science'' (KAKENHI Grant No.~15H05855) from MEXT of Japan.

\end{document}


\title{Topologically nontrivial Andreev bound states: Supplemental Material}
\author{Pasquale Marra}
\author{Muneto Nitta}
\affiliation{Department of Physics, and Research and Education Center for Natural Sciences, Keio University, 4-1-1 Hiyoshi, Yokohama, Kanagawa 223-8521, Japan
}

\begin{abstract}
In the first part of these Supplemental Material, we consider a more general model where we relax some assumptions of the main text, and include the effects of spin-triplet pairing, next-nearest neighbor hopping, and disorder.
Specifically, we show that nontrivial Andreev bound states are robust against these perturbations.
In the second part, we clarify the symmetry properties of the Hamiltonian with respect to the particle-hole symmetry. 
\end{abstract}

\maketitle

\onecolumngrid
\vspace{-10mm}
\section{Generalized Hamiltonian}
\vspace{-2mm}

\begin{figure*}[h]
\centering
(a) \hspace{38mm} (b)\hspace{39mm} (c)\hspace{39mm} (d)\hspace{23mm} \,\\
\includegraphics[height=.27\columnwidth]{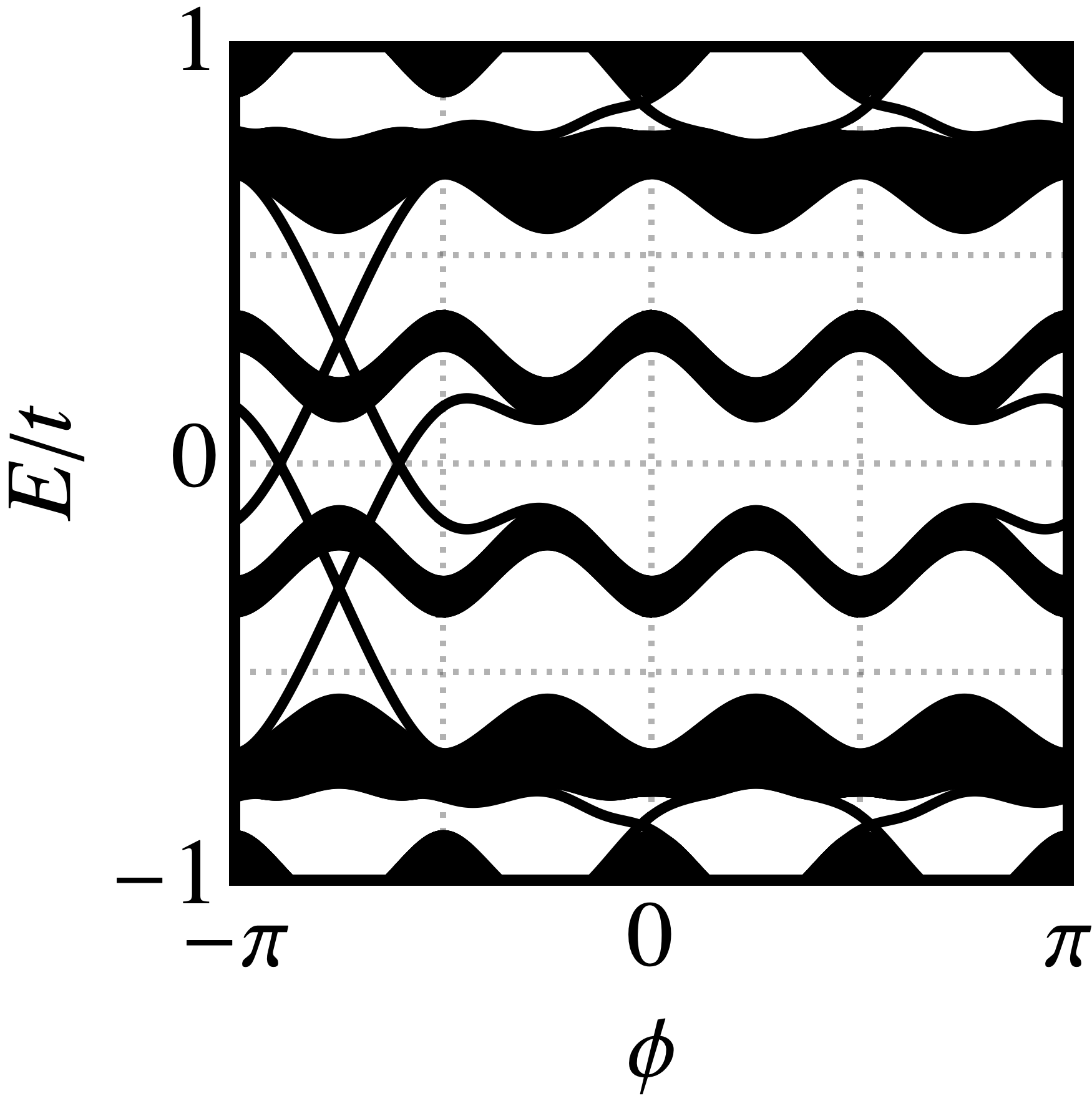}%
\includegraphics[height=.27\columnwidth, trim=20 0 0 0,clip]{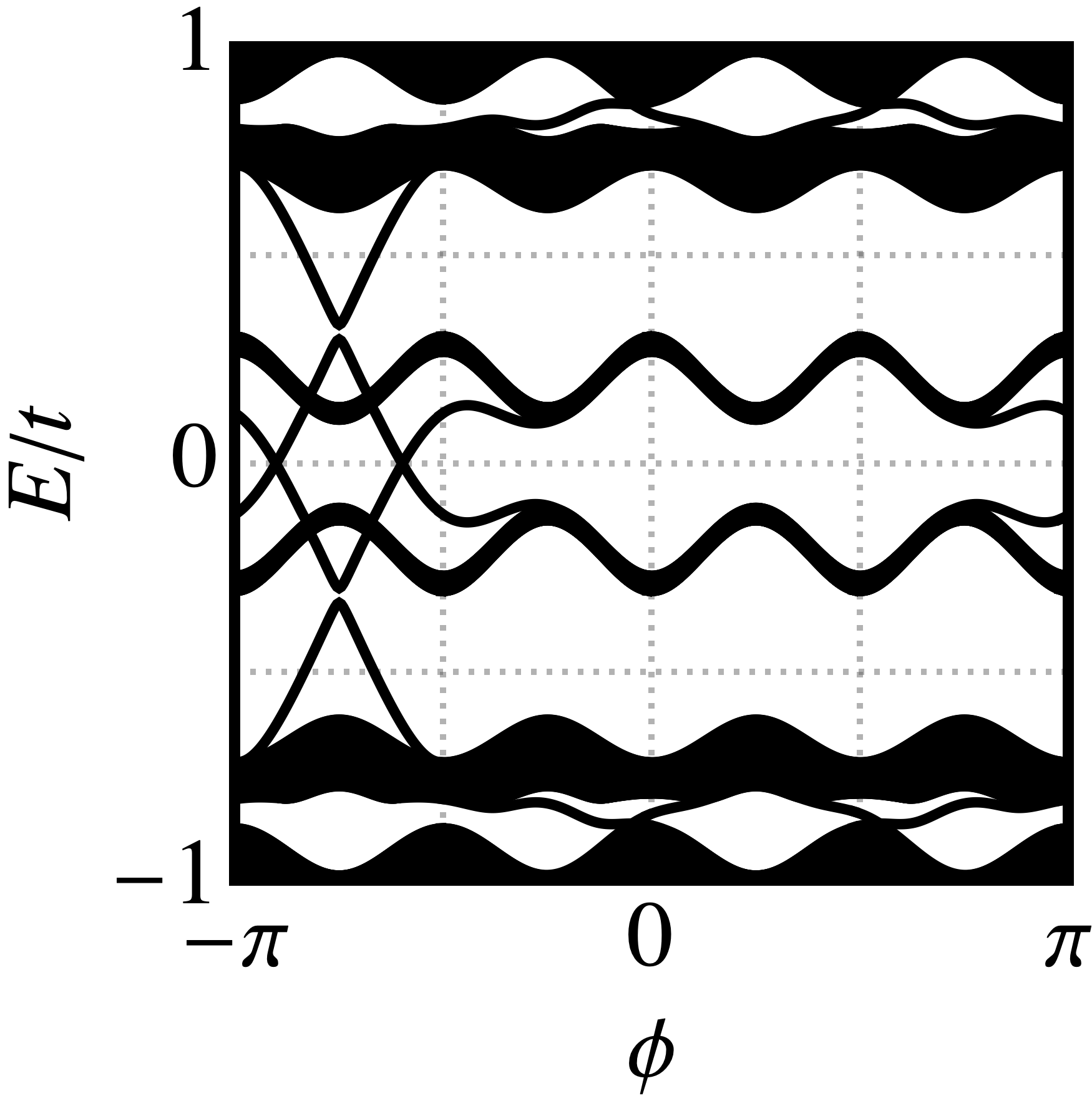}%
\includegraphics[height=.27\columnwidth, trim=20 0 0 0,clip]{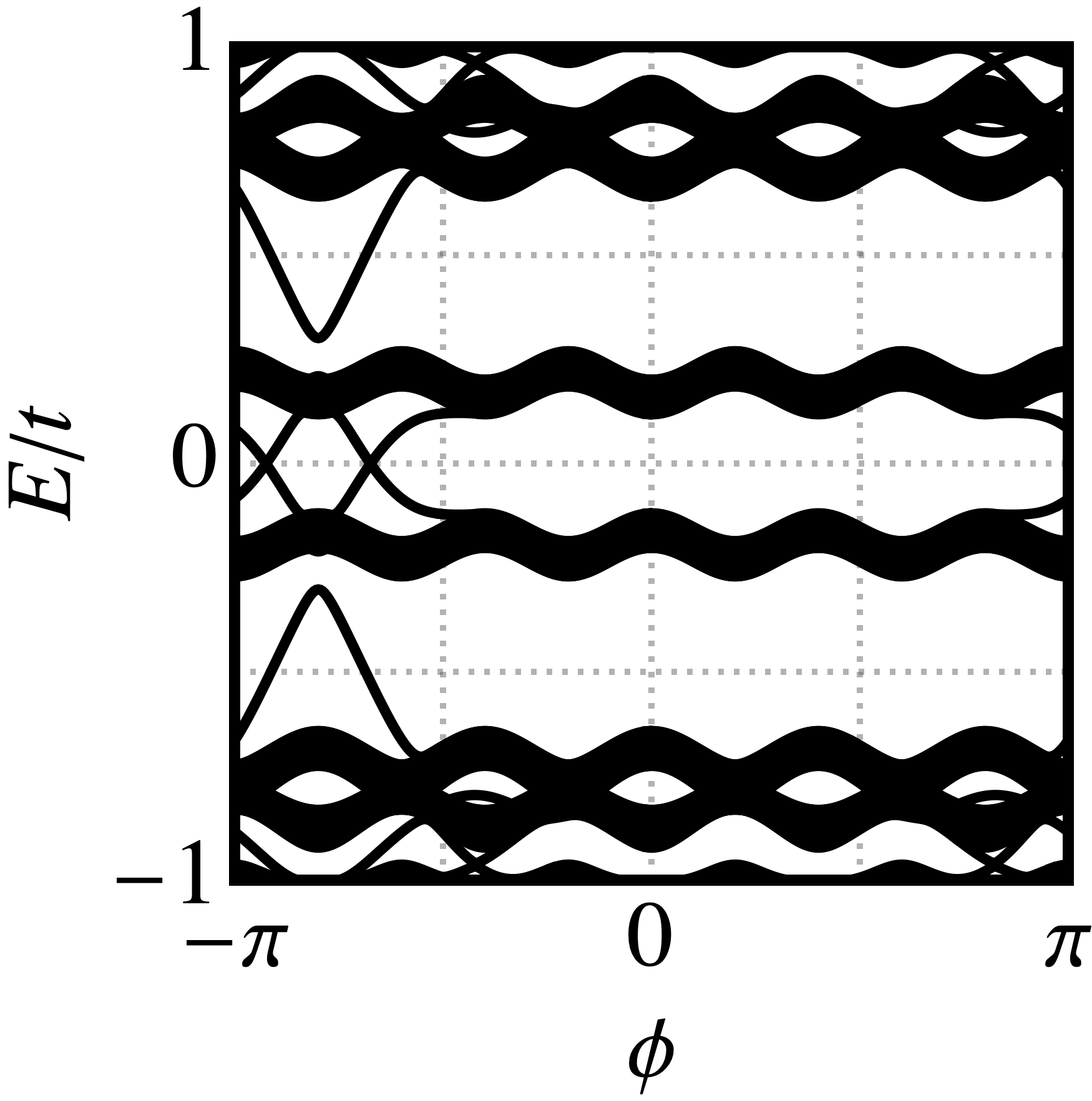}%
\includegraphics[height=.27\columnwidth, trim=20 0 0 0,clip]{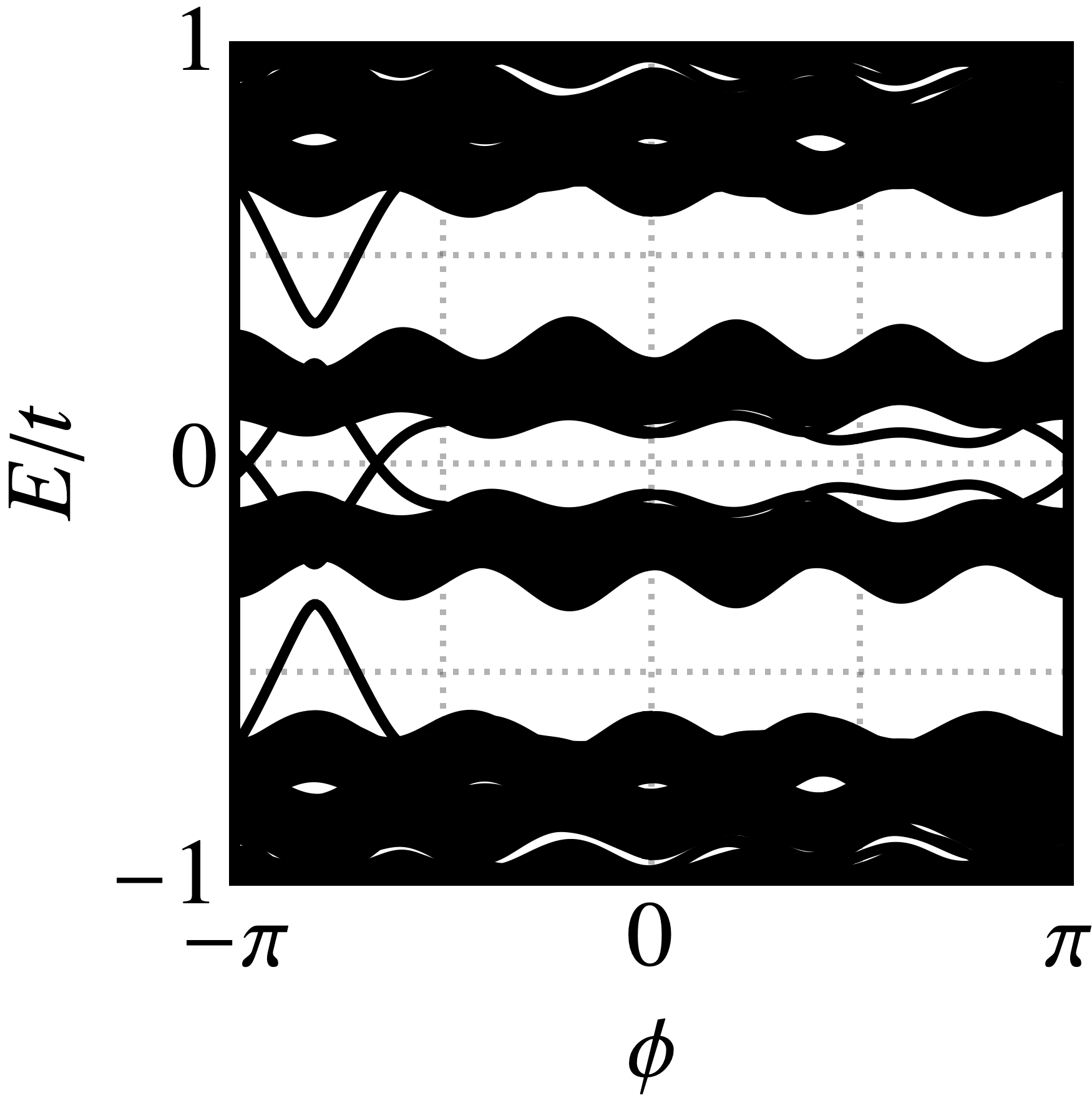}\\[-2mm]
	\caption{%
Energy spectra of the Hamiltonian $H'$ in the case of open boundary conditions, showing the presence of nontrivial ABS below the gap. 
We use  $\mu=-1.2 t$, $\lambda = t/2$,  $\delta b=1.5 t$,  $b=3 t$, $\Delta_s = t/2$ for all figures, and
(a) $\theta = \pi/2$, $t'=0$, $\Delta_p = 0$, 
(b) $\theta = \pi/2$, $t'=0.1t$, $\Delta_p = 0.2t$, 
 (c) $\theta = 2\pi/5$, $t'=0.1t$, $\Delta_p = 0.5t$,
 and (d) same as before but with uncorrelated Gaussian disorder with standard deviation $\sigma=0.05 t$.
}
	\label{fig1}
\end{figure*}

The presence of Rashba spin-orbit coupling induces a spin-triplet pairing, that we can write as
\begin{gather}
H_p=  \sum_{ns} \frac{\Delta_p}2 \left(c^\dagger_{ns} c^\dagger_{n+1,s} - c^\dagger_{n+1,s} c^\dagger_{ns}\right)
\end{gather}
Hence, in order to generalize the Hamiltonian in Eq. 1 of the main text in the presence of spin-triplet pairing, next-nearest neighbor hopping, and disorder, we consider here the following Bogoliubov-de~Gennes Hamiltonian
\begin{gather}
H'=
\frac12\sum_n
\boldsymbol\Psi_{n}^\dag
\!\cdot\!
\begin{bmatrix}
-(\mu+\epsilon_n) \sigma_0
+
\mathbf{b}_n\cdot\boldsymbol{\sigma}
&
\Delta_s\ii\sigma_y\\
-(\Delta_s\ii\sigma_y)^*&
(\mu+\epsilon_n) \sigma_0
-(\mathbf{b}_n\cdot\boldsymbol{\sigma})^*
\end{bmatrix}
\!\cdot\!
\boldsymbol\Psi_{n}
+\nonumber\\
-
\frac12\sum_n
\boldsymbol\Psi_{n}^\dag
\!\cdot\!
\begin{bmatrix}
t\sigma_0
-\lambda \ii\sigma_y
&
\Delta_p \sigma_0\\
- (\Delta_p \sigma_0)^*&
-(t\sigma_0
-\lambda \ii\sigma_y)
\end{bmatrix}
\!\cdot\!
\boldsymbol\Psi_{n+1}
-
\boldsymbol\Psi_{n}^\dag
\!\cdot\!
\begin{bmatrix}
t'\sigma_0
&
0\\
0&
-t'\sigma_0
\end{bmatrix}
\!\cdot\!
\boldsymbol\Psi_{n+2}
+\text{h.~c.}
\label{eq:Hamiltonian}
\end{gather}
Here, 
$
\mathbf{b}_n=\mathbf{b}+\Re(e^{\ii(n \theta+\phi)} \boldsymbol{\delta}\mathbf{b})
$ 
is the onsite Zeeman field, 
$\Delta_s$ and $\Delta_p$ are respectively the spin-singlet and spin-triplet superconducting pairing, 
$t$ and $t'$ are the nearest and next-nearest neighbor hopping, 
$\lambda$ is the intrinsic spin-orbit coupling, 
and $\epsilon_n$ is the onsite uncorrelated Gaussian disorder.
We moreover relax the assumption on $\theta$ and $k_\mathrm{F}$ of the main text and we assume in general $\theta\neq 2 k_\mathrm{F}$.
\Cref{fig1} shows the energy spectra of the Hamiltonian in \cref{eq:Hamiltonian} with open boundary conditions calculated for various regimes.
Nontrivial ABS are robust against the perturbations considered here.

\vspace{-3mm}
\section{PH Symmetry of the Hamiltonian}
\vspace{-2mm}

The Hamiltonian in Eq. 2 of the main text is invariant with respect to the PH symmetry operator $\Xi$, which we will define below.
This symmetry operator $\Xi$ acts as $H(k,\phi)\to-H(-k,\phi)$, in contrast with the case of, e.g., topological superconductors in 2 real (non-synthetic) dimensions, where one has $H(k_x,k_y)\to-H(-k_x,-k_y)$.
This is a consequence of the fact that in model considered, the superconducting  term couples fermions with opposite momenta and same phase $(k,\phi)$ and $(-k,\phi)$, in contrast with the case of superconductors in 2 real (non-synthetic) dimensions, where the superconducting term couples fermions with opposite momenta $(k_x,k_y)$ and $(-k_x,-k_y)$. 
Physically, this corresponds to the fact that the variable $\phi$ is time-like and not spatial-like, and the superconducting pairing considered here is frequency-even.
To illustrate this point, we first start with a general example and we then address the case of the model considered in the main text. 

Let us consider a spin-singlet $s$-wave superconductor in 1 dimension with an additional synthetic dimension $\phi$.
We assume that the synthetic dimension $\phi$ is time-like, in the sense that, at each instant in time, the system is described by a well-defined value of the phase $\phi$.
In this case the superconducting  term couples fermions with opposite momenta and spin, and with the \emph{same} phase $\phi$, i.e.,
$
\sum_{{ k} } \Delta 
(
c^\dag_{{ k,\phi} \up} c^\dag_{-{ k},\phi \down}
-
c^\dag_{{ k,\phi} \down} c^\dag_{-{ k},\phi \up}
)
$,
and the Bogoliubov-de~Gennes Hamiltonian reads
\begin{equation}\label{canonical}
H=\sum_{k}
\boldsymbol\Psi_{{ k,\phi}}^\dag
\!\cdot\!
H({ k},\phi)
\!\cdot\!
\boldsymbol\Psi_{{ k,\phi}},
\qquad\text{with }
H({ k},\phi)=
\begin{bmatrix}
h({ k},\phi)
&
\ii\Delta\sigma_y\\
(\ii\Delta\sigma_y)^\dag&
-h(-{ k},\phi)^*
\end{bmatrix},
\end{equation}
where $\boldsymbol\Psi_{{ k,\phi}}^\dag=[c_{{ k,\phi} \up}^\dag c_{{ k,\phi} \down}^\dag c_{-{ k},\phi \up} c_{-{ k},\phi \down} ]$.
Thus, the particle and hole sectors are respectively $h(k,\phi)$ and $-h(-k,\phi)^*$.
In this case the PH symmetry operator is $\Xi=\tau_x K$, where $K$ is the complex conjugation.
This operator satisfies $\Xi^2=\id$, and
\begin{equation}\label{PHphase}
\Xi H({ k},\phi) \Xi^{-1}=- H(-{ k},\phi).
\end{equation}
Therefore, the  PH symmetry acts like $H({ k},\phi) \to - H(-{ k},\phi)$.
As a counterexample, one can consider the case of a spin-singlet $s$-wave superconductor in 2 real (non-synthetic) dimensions.
In this case the superconducting  term couples fermions with opposite momenta and spin, i.e.,
$
\sum_{{\mathbf k} } \Delta 
(
c^\dag_{{\mathbf k} \up} c^\dag_{-{\mathbf k} \down}
-
c^\dag_{{\mathbf k} \down} c^\dag_{-{\mathbf k} \up}
)
$, and the Bogoliubov-de~Gennes Hamiltonian reads
\begin{equation}
 H=\sum_{\mathbf k}
\boldsymbol\Psi_{{\mathbf k}}^\dag
\!\cdot\!
H({\mathbf k})
\!\cdot\!
\boldsymbol\Psi_{{\mathbf k}}
\qquad
H({\mathbf k})=
\begin{bmatrix}
h({\mathbf k})
&
\ii\Delta\sigma_y\\
(\ii\Delta\sigma_y)^\dag&
-h(-{\mathbf k})^*
\end{bmatrix}
\end{equation}
where $\boldsymbol\Psi_{{\mathbf k}}^\dag=[c_{{\mathbf k} \up}^\dag c_{{\mathbf k} \down}^\dag c_{-{\mathbf k} \up} c_{-{\mathbf k} \down} ]$.
The particle and hole sectors are respectively $h(k_x,k_y)$ and $-h(-k_x,-k_y)^*$, and
the PH symmetry 
$\Xi=\tau_x K$ is such that
\begin{equation}
\Xi H({\mathbf k}) \Xi^{-1}=- H(-{\mathbf k}).
\end{equation}
Therefore, the  PH symmetry acts like $H(k_x,k_y) \to - H(- k_x,-k_y)$.

The definition of PH symmetry for the Hamiltonian in Eq. 2 in the main text is slightly more involved, because this Hamiltonian is not diagonal with respect to the momentum $k$, but contains terms which couple fermionic states at different momenta $k$ and $k+n\theta$.
This Hamiltonian can be written explicitly in matrix form as
\begin{equation}\label{Hfull}
 H=
\frac12
\sum_k
\boldsymbol\Psi_{{ k}}^\dag
\!\cdot\!
H(k,\phi)
\!\cdot\!
\boldsymbol\Psi_{{ k}}
,
\qquad\text{with }
H(k,\phi)=
\begin{bmatrix}
 H_p(k,\phi) & \ii\Delta\sigma_y\id_q\\
 (\ii\Delta\sigma_y \id_q)^\dag&   H_h(k,\phi)
\end{bmatrix},
\end{equation}
where is
$\id_q$ the identity matrix of rank $q$, and where the particle and hole sectors are given by 
\begin{equation}\label{Hph}
 H_p(k,\phi)=
\begin{bmatrix}
h(k) & w(\phi) &\\
w(\phi)^\dag & h(k+\theta) &\\[-2.5mm]
&&\!\!\ddots\!\!\\[-1.5mm]
 && &h(k+(q-1)\theta) \\
\end{bmatrix},
\quad 
 H_h(k,\phi)=
-
\begin{bmatrix}
h(-k) & w(\phi) &\\
w(\phi)^\dag & h(-k-\theta) &\\[-2.5mm]
&&\!\!\ddots\!\!\\[-1.5mm]
 && &h(-k-(q-1)\theta) \\
\end{bmatrix}^*,
\end{equation}
 with 
$h(k)=
\mathbf{b}\!\cdot\!\boldsymbol{\sigma}
-(\mu+2t\cos{k})\sigma_0
+2\lambda\sin{k}\,\sigma_y
$, and
$w(\phi)=\frac{1}2 e^{\ii\phi}\boldsymbol{\delta}\mathbf{b}\cdot\boldsymbol{\sigma}$.
The superconducting pairing couples fermions with opposite momenta, but with the same phase $\phi$.
The Nambu vector is   
\begin{equation}
\boldsymbol\Psi_{{ k}}^\dag=
[
c_{{ k} \up}^\dag , c_{{ k} \down}^\dag ,
c_{{ k+\theta} \up}^\dag , c_{{ k+\theta} \down}^\dag \,
\dots\,
c_{{ k+(q-1)\theta} \up}^\dag , c_{{ k+(q-1)\theta} \down}^\dag ,
c_{{- k} \up} , c_{{ -k} \down} ,
c_{{ -k-\theta} \up} , c_{{ -k-\theta} \down} \,
\dots\,
c_{{ -k-(q-1)\theta} \up}^\dag , c_{{- k-(q-1)\theta} \down} 
].
\end{equation}
The Hamiltonian has the advantage that the superconducting term $\ii\Delta\sigma_y\id_q$ is diagonal in the momentum basis.
However it has the disadvantage that $H_h(k,\phi)\neq-H_p(-k,\phi)$.
To write this Bogoliubov-de Gennes Hamiltonian in the canonical form, one needs to reorder the Nambu basis with a unitary transformation.
To do so, one can define the unitary operator $V_q$ as the operator whose action is to reorder the terms of the Hamiltonian hole sector $H_h(k,\phi)$ such that
\begin{equation}\label{dio1}
V_q
\cdot
\begin{bmatrix}
h(-k) & w(\phi) &\\
w(\phi)^\dag & h(-k-\theta) &\\[-2.5mm]
&&\ddots\\[-1.5mm]
 && &h(-k-(q-1)\theta) 
\end{bmatrix}
\cdot
V_q^\dag
=
\begin{bmatrix}
h(-k) & w(\phi) &\\
w(\phi)^\dag & h(-k-(q-1)\theta) &\\[-2.5mm]
&&\ddots\\[-1.5mm]
 && &h(-k-\theta) 
\end{bmatrix}
.
\end{equation}
Using that $h(k+2\pi)=h(k)$ and that $-(q-n)\theta = n\theta -2\pi p$ one has that
\begin{equation}\label{dio2}
\begin{bmatrix}
h(-k) & w(\phi) &\\
w(\phi)^\dag & h(-k-(q-1)\theta) &\\[-2.5mm]
&&\ddots\\[-1.5mm]
 && &h(-k-\theta) 
\end{bmatrix}
=
\begin{bmatrix}
h(-k) & w(\phi) &\\
w(\phi)^\dag & h(-k+\theta) &\\[-2.5mm]
&&\ddots\\[-1.5mm]
 && &h(-k+(q-1)\theta) \\
\end{bmatrix}
.
\end{equation}
\Cref{Hph,dio1,dio2} give
\begin{equation}
V_q H_h(k,\phi) V_q^\dag=-H_p(-k,\phi)^*
\end{equation}
and therefore one can write
\begin{equation}\label{Hfull2}
H(k,\phi)=
\begin{bmatrix}
 H_p(k,\phi) & \ii\Delta\sigma_y\id_q\\
 (\ii\Delta\sigma_y\id_q)^\dag&   H_h(k,\phi)
\end{bmatrix}
=
\begin{bmatrix}
 H_p(k,\phi) & \ii\Delta\sigma_y\id_q\\
 (\ii\Delta\sigma_y\id_q)^\dag&  
- V_q^\dag  H_p(-k,\phi)^*   V_q
\end{bmatrix}
.
\end{equation}
Finally, we rewrite the Hamiltoinian in \cref{Hfull} in the new basis, as
\begin{gather}
\bar H(k,\phi)
=
\begin{bmatrix}
1&0\\
0&V
\end{bmatrix}
\cdot
\begin{bmatrix}
 H_p(k,\phi) & \ii\Delta\sigma_y\id_q\\
 (\ii\Delta\sigma_y\id_q)^\dag&  
- V_q^\dag  H_p(-k,\phi)^*   V_q
\end{bmatrix}
\cdot
\begin{bmatrix}
1&0\\
0&V^\dag
\end{bmatrix}
=
\begin{bmatrix}
 H_p(k,\phi) &( \ii\Delta\sigma_y\id_q) V^\dag\\
 V (\ii\Delta\sigma_y\id_q)^\dag &  
-   H_p(-k,\phi)^*   
\end{bmatrix}
,
\end{gather}
which gives
\begin{equation}\label{HfullS}
 H=
\frac12
\sum_k
\boldsymbol\Psi_{{ k}}^\dag
\!\cdot\!
\bar H(k,\phi)
\!\cdot\!
\boldsymbol\Psi_{{ k}}
,
\qquad\text{with }
\bar H(k,\phi)=
\begin{bmatrix}
 H_p(k,\phi) &\bar \Delta \\
 \bar \Delta^\dag&   H_p(-k,\phi)^*
\end{bmatrix},
\end{equation}
where $\bar \Delta=(\ii\Delta\sigma_y\id_q V^\dag) $.
Hence, by reordering the momentum basis with the unitary operator $V_q$, the Hamiltonian is in a canonical form, analogous to the case in \cref{canonical}.
As before, the antiunitary symmetry operator $\Xi$ in momentum space is
\begin{equation}\label{PHC}
\Xi= \tau_x K ,
\end{equation}
satisfies $\Xi^2=\id$, and also
\begin{equation}
\Xi \bar H(k,\phi) \Xi^{-1} =-\bar  H(-k,\phi).
\end{equation}

We have shown that the Hamiltonian in Eq. 2 of the main text is invariant with respect to the PH symmetry operator $\Xi$ defined in \cref{PHC}.
As we anticipated, in this case the symmetry operator $\Xi$ acts as $H(k,\phi)\to -H(-k,\phi)$, in contrast with the case of, e.g., topological superconductors in 2 non-synthetic dimensions, where $H(k_x,k_y)\to-H(-k_x,-k_y)$.
We stress that this is a consequence of the fact that in the model considered, the superconducting term couples fermions with opposite momenta but \emph{same} phase $(k,\phi)$ and $(-k,\phi)$, in contrast with the case of superconductors in 2 real (non-synthetic) dimensions, where the superconducting term couples fermions with $(k_x,k_y)$ and $(-k_x,-k_y)$.

\begin{figure*}[h]
\centering
(a) \hspace{38mm} (b) \hspace{23mm} \,\\
\includegraphics[height=.27\columnwidth]{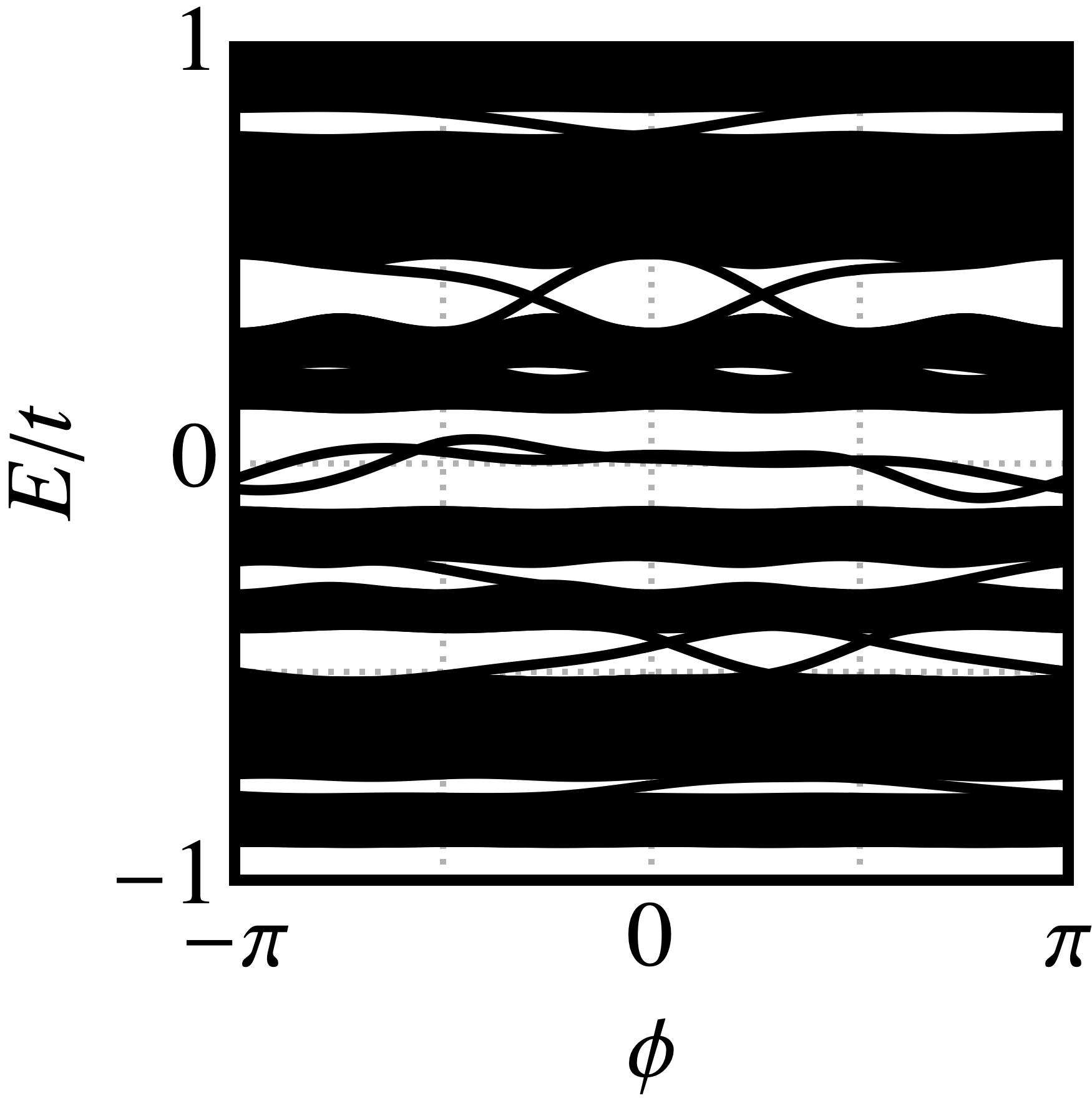}%
\includegraphics[height=.27\columnwidth, trim=20 0 0 0,clip]{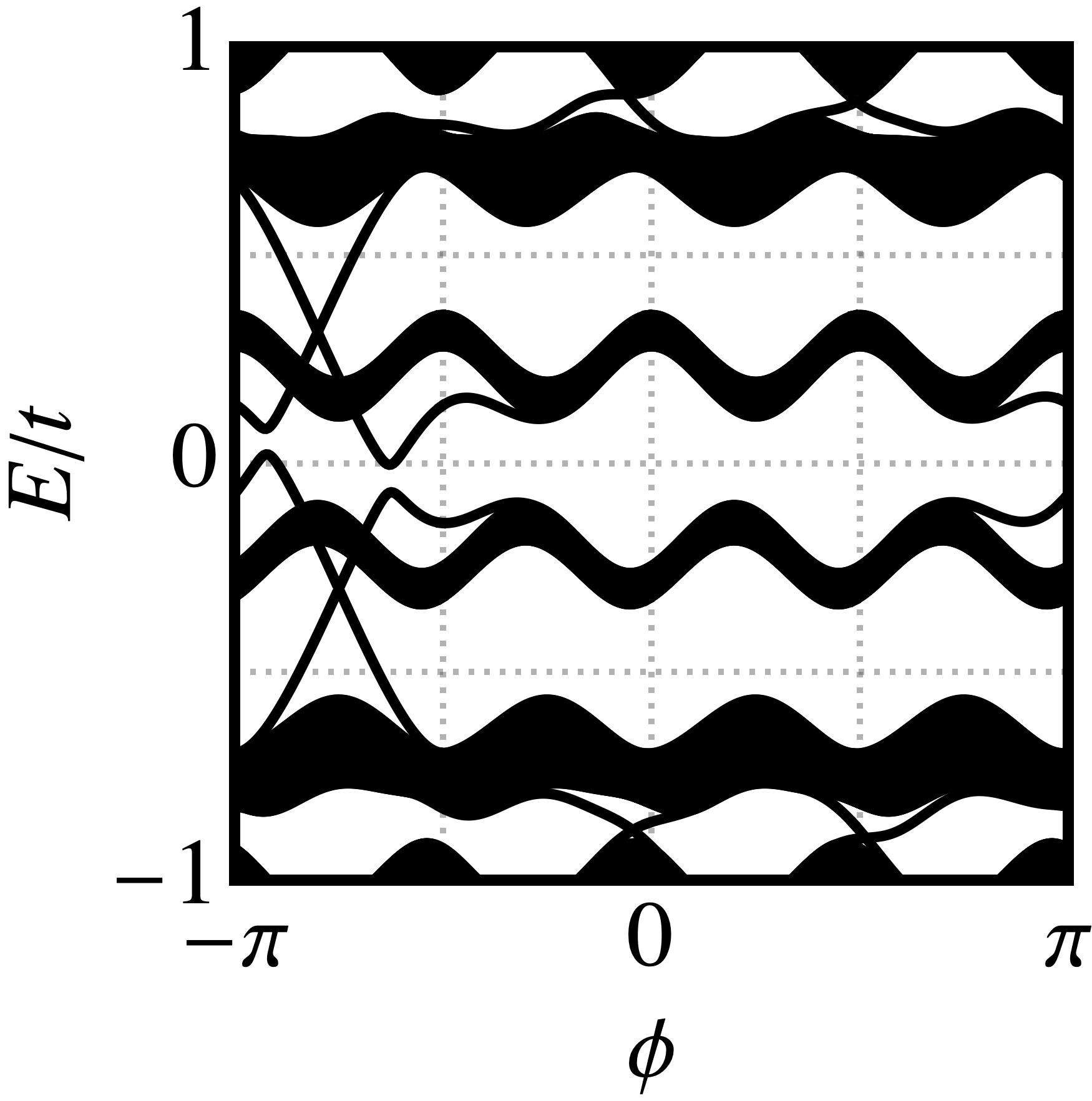}\\[-2mm]
	\caption{%
Energy spectra of the Hamiltonian $H+H_\mathrm{SB}$ for $\eta=t/4$.
All parameters are as in Fig. 2 (b-c) of the main text.
MBS and nontrivial ABS are not robust against perturbations PH symmetry-breaking term $H_\mathrm{SB}$.
}
	\label{fig2}
\end{figure*}

Finally, we verify numerically that the nontrivial ABS are protected by the PH symmetry.
In order to do so, we introduce a PH symmetry-breaking term
\begin{equation}
H_\mathrm{SB}=
\eta \sum_n \boldsymbol\Psi_{n}^\dag \tau_0 \sigma_x \boldsymbol\Psi_{n}
=
\eta \sum_k \boldsymbol\Psi_{k}^\dag \tau_0 \sigma_x \boldsymbol\Psi_{k}
,
\end{equation}
and calculate the energy spectra for the Hamiltonian $H+H_\mathrm{SB}$ with open boundary conditions. 
\Cref{fig2} shows that both MBS and nontrivial ABS are not robust against the PH symmetry-breaking term $H_\mathrm{SB}$.

